\documentclass[seceq]{ptptex}

\usepackage{graphicx}

\title{
Recent Heavy Flavor Results at RHIC%
}

\author{
Wenqin \textsc{Xu}%
}

\inst{
Department of Physics and Astronomy, University of California, Los Angeles}

\abst{
We summarize the recent experimental results of heavy flavor physics from the Relativistic Heavy Ion Collider (RHIC) in Brookhaven National Lab (BNL) at Long Island, New York, USA. We will discuss the directly reconstructed open charm mesons as well as electrons from heavy flavor hadron decays. The charm and bottom quark production cross-sections have also been measured. We will also discuss $J/\psi$ and $\Upsilon$ states in $p+p$ and heavy ion collisions. The studies described here were carried out and reported by the STAR and PHENIX collaborations at RHIC.
}

\begin{document}

\maketitle

\section{Introduction}
The hot and strongly coupled QCD matter created in the relativistic Heavy Ion collisions, namely Quark-Gluon Plasma (QGP), has been studied extensively by various methods with different probes. For example, it is experimentally found that the elliptic flow ($v_2$)~\cite{Art98a} of bulk soft particles 
can be described by hydrodynamics with a small shear viscosity over entropy density ratio ($\eta /s$), and attempts have been made to extract this quantity~\cite{Shen11a}. Another entirely different type of probes are heavy flavor hadrons, either open or hidden. Due to the large masses of heavy quarks, the energy scales involved are often much larger than the QCD scale $\Lambda$, so that the perturbative QCD (pQCD) calculations are applicable.  As a result, the heavy flavor probes, along with other rare probes such as high transverse momentum ($p_T$) particles and jets, provide possible bridges to quantitative understandings of QGP. In these proceedings, we will focus on recent measurements made with heavy flavor probes by the PHENIX and STAR collaborations at RHIC.

\section{Open heavy flavor probes}
\begin{figure}[tb]
\centering
\begin{tabular}{cc}  
  \includegraphics[width=0.24 \linewidth]{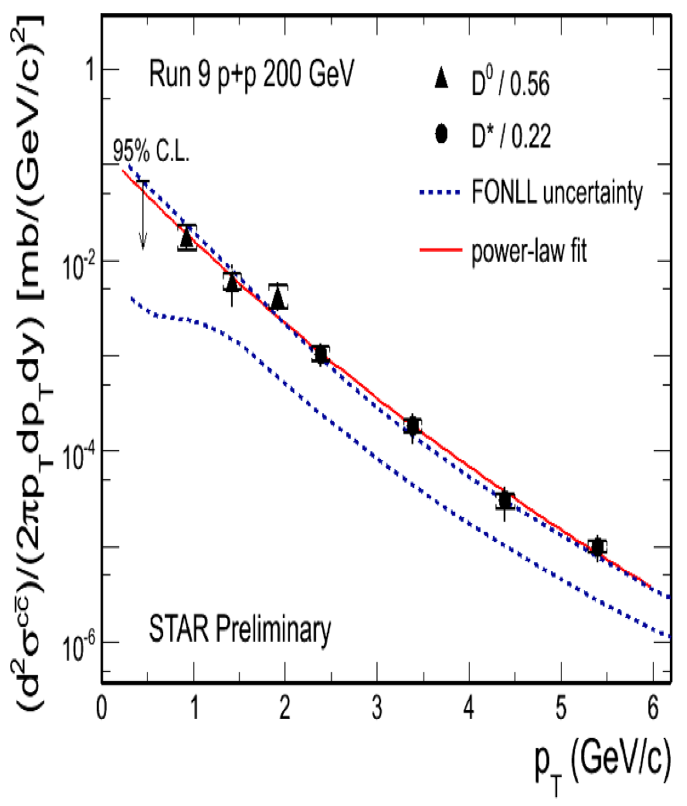}
 \includegraphics[width=0.5 \linewidth]{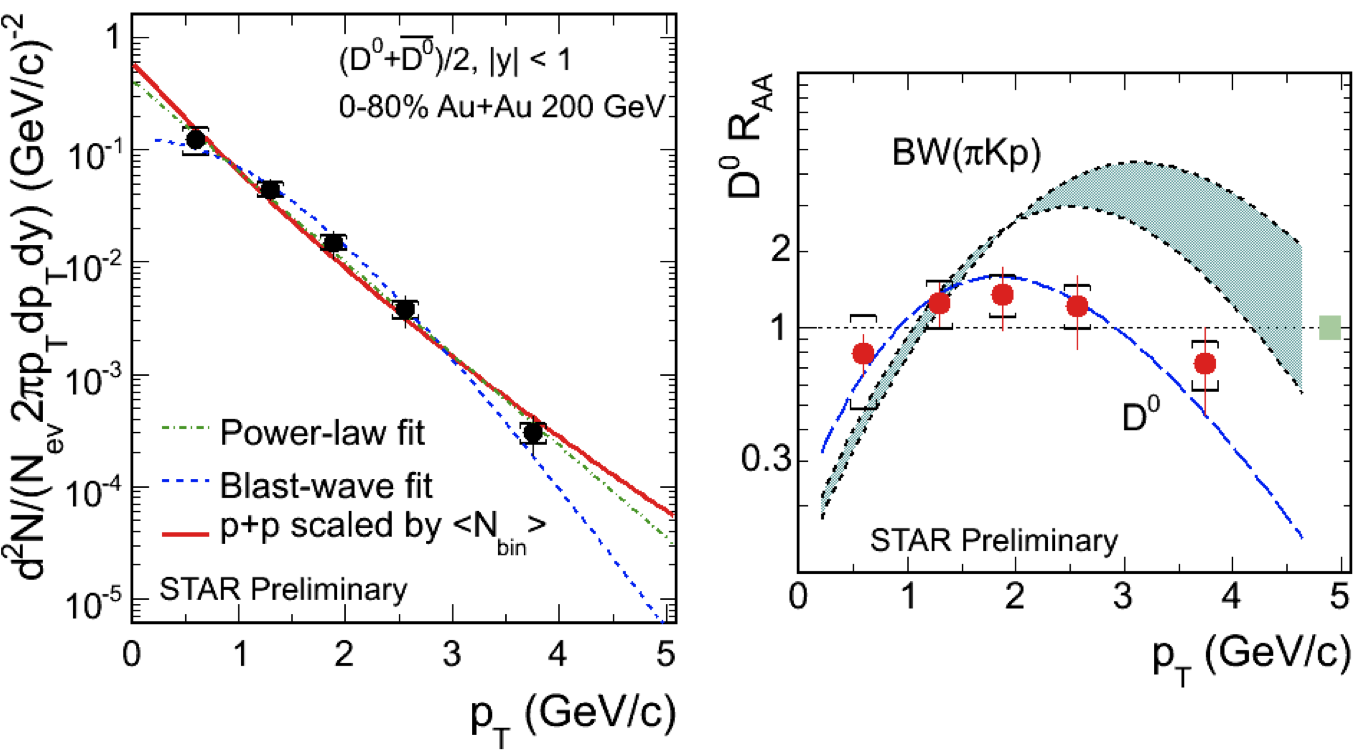}
\end{tabular}
\caption{Left: $p_T$ dependence of charm quark production cross-section based on directly reconstructed charm mesons (both $D^{0}$ and $D^{*}$)  in $p+p$ collisions at $\sqrt{s}= 200$~GeV.  The FONLL ~\cite{Matteo05a} calculations are also shown as dashed curves. Middle: $p_T$ spectrum of $D^{0}$ in 0-80\% central events of $Au+Au$ collisions at $\sqrt{s_{NN}}= 200$~GeV. Right: $D^{0}~R_{AA}$ in 0-80\% central events of $Au+Au$ collisions. All the plots are adapted from reference~\cite{Zhang11a}}
\label{fig:Figpp}
\end{figure}

The heavy quarks are dominantly produced in the hard scatterings in the inital stage of heavy ion collisions, and pQCD calculations should be applicable. This has be checked and confirmed by recent measurements in $p+p$ collisions at 200GeV. The preliminary result of $p_T$ spectrum of charm quark production in $p+p$ collisions, based on directly reconstructed charm mesons (both $D^{0}$ and $D^{*}$), has been reported by STAR~\cite{Zhang11a}, as shown in the left plot of Figure~\ref{fig:Figpp}. The $p_T$ spectrum of Non-photonic Electrons (NPE) or Heavy-Flavor Electrons, {\it i.e.} the electron daughters of the semi-leptonic decays of open heavy flavor hadrons, has also recently been published by PHENIX~\cite{PHENIX10a} and STAR~\cite{STAR11a}. For both measurements in $p+p$ collisions, Fixed-Order Next-to-Leading Logarithm (FONLL) calculations~\cite{Matteo05a} are found to be consistent with the data within the uncertainty. Generally speaking, FONLL upper boundary is more in line with the data points, except in the case of electrons solely from bottom decays, where the central values of the same FONLL calculations are more consistent with the data~\cite{STAR11a}. 

The experimental details can be found in~\cite{STAR05a,Zhang11a} for direct charm meson reconstructions and charm quark production cross section extraction and in~\cite{PHENIX10a, STAR11a} for NPE analyses. Here we will just brief one key experimental upgrade in STAR. The direct charm meson reconstruction requires particle identification (PID) for $\pi^{\pm}$ and $K^{\pm}$, which was done in STAR solely based on track energy loss in the Time Projection Chamber (TPC)~\cite{STAR03a} gas in the past. The STAR Time-Of-Flight detector(TOF)~\cite{TOF04a} is a key upgrade that was successfully accomplished in 2010, and it extends the STAR PID ability to above 1GeV/c, largely improving many measurements at STAR, including $D^{0}$ and $D^{*}$ reconstructions.

If there is no significant gluon shadowing in the relevant kinematics region for charm quarks at RHIC, then the total charm quark production per nucleon-nucleon collision($\sigma^{NN}_{c\bar c}$) should be the same in both $p+p$ and heavy ion collisions. The preliminary STAR results at mid-rapidity (${\sigma^{NN}_{c\bar c}\over dy}\mid_{y=0}$), which are $202\pm56(stat.)\pm40(sys.)\pm20(norm.)\mu$b in $p+p$ and $186\pm22(stat.)\pm30(sys.)\pm18(norm.)\mu$b in $Au+Au$~\cite{Zhang11a}, indeed don't show a dependence on number of binary collisions ($N_{bin}$) calculated by Glauber model~\cite{Miller07a}. The $p_T$ spectrum of $D^{0}$ in 0-80\% central events of $Au+Au$ collisions at $\sqrt{s_{NN}}= 200$~GeV has also been measured (the middle plot in Figure~\ref{fig:Figpp}) and compared to that in $p+p$ collisions scaled by proper $N_{bin}$ to get the $D^0$ nuclear modification factor($R_{AA}$) (the right plot in Figure~\ref{fig:Figpp}). No suppression is seen for $D^0$ with $p_T<$~3GeV/c. It's worth pointing out that in the semi-leptonic decays of heavy flavor hadrons, say $D^0$, the electron daughters often have much smaller $p_T$, {\it e.g.} around half of parent $D^0$ $p_T$. As a result, charm decay electrons and $D^0$ with same $p_T$ usually represent charm quarks with different $p_T$.

To access the higher $p_T$ regime of heavy flavor hadrons/quarks, one can effectively trigger on non-photonic electrons with calorimeters. Both STAR and PHENIX have found the NPE suppression in central $Au+Au$ collisions is as much as that of light particles~\cite{STAR11b, PHENIX07a}, contradicting to long standing pQCD predications~\cite{Yu01a}. To solve this puzzle, measurements with higher precisions in both $p+p$ and $Au+Au$ collisions are required. PHENIX has already published such measurements~\cite{PHENIX10a} and STAR, formerly constrained by large photonic electron background due to large material budget, is also making significant progresses. STAR has published much improved NPE measurement in $p+p$~\cite{STAR11a} and they are working on $Au+Au$ collisions~\cite{Wenqin11a}.

It is important to experimentally differentiate bottom and charm contributions to NPE yield, so that definitive comparisons with theoretical calculations could be possible for either species. Both experiments attack this problem considering correlations between NPE and hadrons. STAR focuses on the azimuthal correlations~\cite{STAR10a}, and PHENIX studies the invariant mass correlations~\cite{PHENIX09a}. These correlations originate from the production and decay of heavy quarks/hadrons, and they are different for charm and bottom. Currently, these methods have only been applied in $p+p$ collisions. Furthermore, STAR measured the spectra for charm and bottom decay electrons separately in $p+p$ collisions~\cite{STAR11a}. STAR normalized PYTHIA model calculations to the measured bottom decay electron spectrum and extracted the bottom production cross section in $p+p$ collisions at $\sqrt{s}= 200$~GeV to be between $1.34$ and $1.83~\mu $b~\cite{STAR11a}. It was also concluded that bottom hadron production is suppressed at high $p_T$ in $Au+Au$ collisions~\cite{STAR10a} . Due to limited pages, we skip the details and other  NPE measurements, such as elliptic flow and NPE-hadron azimuthal correlations, and refer to the literature~\cite{Wenqin11b,PHENIX10b} 

\section{Heavy quarkonium}
The suppression of $J/\psi$ due to color-screening induced disassociations was proposed as the signature of QGP~\cite{Matsui86a}. However, recently it is realized that many processes, such as regeneration from charm quarks in the QCD medium, could contribute to $J/\psi$ productions, which make the picture much more complicated. Also, even the direct $J/\psi$ production itself is non-trival for modeling and constraints from data are valuable. 

\begin{figure}[tb]
\centering
\begin{tabular}{cc}  
  \includegraphics[width=0.35 \linewidth]{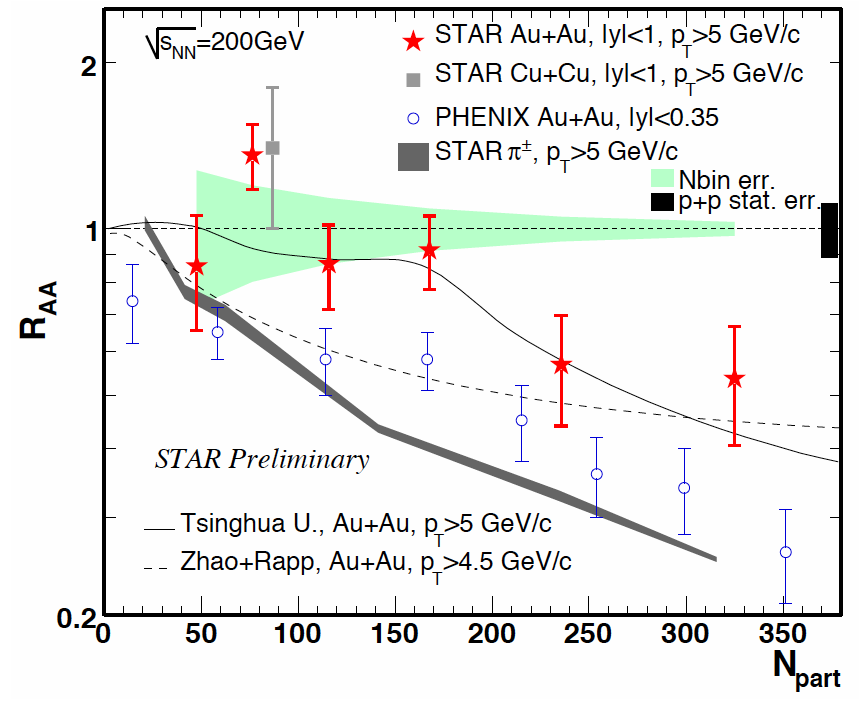}
  \includegraphics[width=0.4 \linewidth]{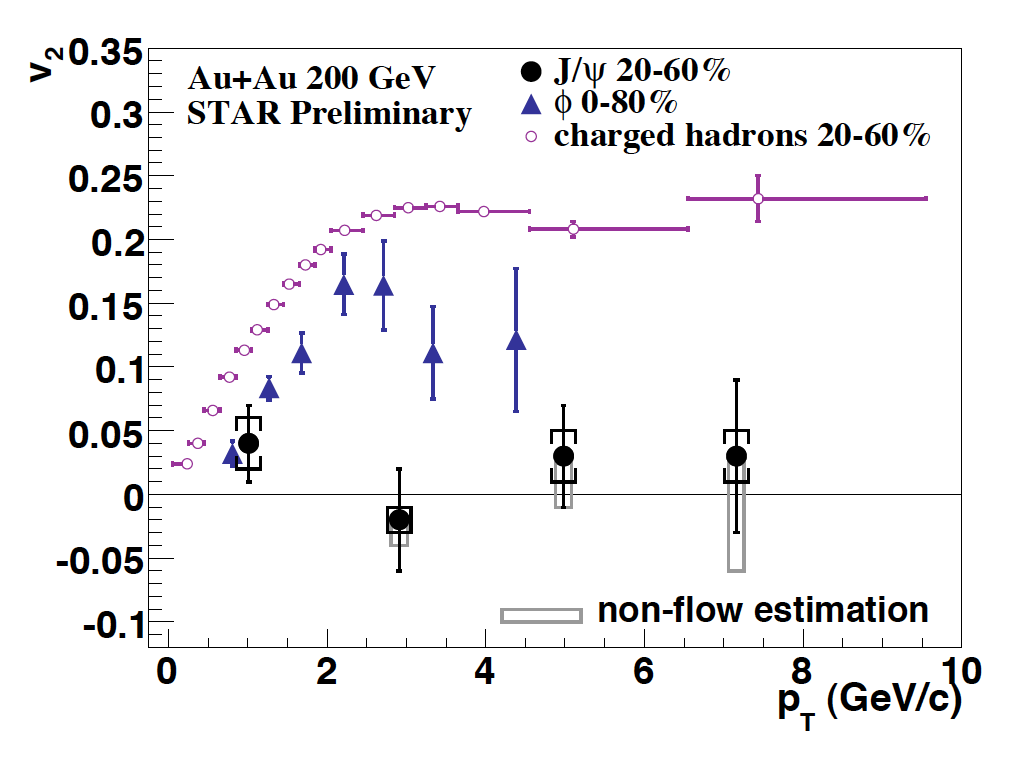}
\end{tabular}
\caption{Left: $N_{part}$ dependence of $J/\psi$ $R_{AA}$.  PHENIX data points are for $J/\psi$ with $0<p_T<5$~GeV/c and STAR for $p_T>5$~GeV/c. Right: $J/\psi$ elliptic flow in $20-60\%$ central events. Both plots are adapted from reference~\cite{Zebo11a}}
\label{fig:FigJpsi}
\end{figure}

At RHIC, $J/\psi$ productions are measured in both $p+p$ and $Au+Au$ collisions. While the PHENIX results cover the low $p_T$ range with high precision, the STAR preliminary results extend the spectra to $p_T$ around~9~GeV/c, and they are in good agreement with each other in the overlap regions. The nuclear modification factor $R_{AA}$ for $J/\psi$ shows an increase from low $p_T$ to high $p_T$. Shown in the left plot of Figure~\ref{fig:FigJpsi}, the $p_T$ integrated $R_{AA}$ for $J/\psi$ is consistent with unity for (semi-)peripheral $Au+Au$ collisions and is smaller than 1 for central collisions. The suppression of higher states could play a big role in this. Shown in the right plot of Figure~\ref{fig:FigJpsi}, the STAR preliminary results of $J/\psi$ elliptic flow ($v_2$) is consistent with zero for $p_T$ above 3~GeV/c, and may be finite at $p_T$ around 1~GeV/c. More investigations are required to understand the picture completely.

PHENIX has found that $J/\psi$ is more suppressed at forward rapidity~\cite{PHENIX11a}, suggesting cold nuclear matter effects are playing significant roles. More detailed studies have been carried out by PHENIX. For example, studies in $d+Au$ and $p+p$ collisions are reported~\cite{PHENIX10c} and they may indicate the effects of gluon saturation.

STAR has recently made the first $\Upsilon$ $R_{AA}$ measurement at RHIC and the preliminary result~\cite{Rosi11a}, although with large uncertainties, suggests the sum of three $\Upsilon$ S-states is indeed suppressed in the most central $Au+Au$ collisions. More statistics and better differentiations between the three states are needed to gain quantitative conclusions.

In summary, RHIC heavy flavor program has been very rich with some conclusions being drawn and more novel phenomena demanding understandings. With newly finished/planned experimental upgrades and emerging physical insights, RHIC heavy flavor program will remain as an excellent platform for more precise and quantitive data-model comparisons.

We would like to thank Z. Tang and Y. Zhang for their helpful discussions.


%

\end{document}